\documentstyle[prl,aps,twocolumn]{revtex}

\begin{document}


\title{Electronic and magnetic structures of Sr$_2$FeMoO$_6$}

\author{Sugata Ray,$^{\star}$ Ashwani Kumar,$^{\star}$ D. D.
Sarma,$^{\star,}$\cite{jnc}
R. Cimino,$^{\S}$ S. Turchini,$^{\|}$ S. Zennaro$^{\P}$ and N. Zema$^{\sharp}$}

\address{$^{\star}$Solid State and Structural Chemistry Unit, Indian Institute
of
Science, Bangalore 560~012, INDIA \\ $^{\S}$Istituto Nazionale di
Fisica Nucleare - Laboratori Nazionali di Frascati, ITALY \\
$^{\|}$Istituto di Chimica
dei Materiali, CNR- Area della Ricerca di Montelibretti - Roma, ITALY\\
$^{\P}$Istituto di Struttura della Materia, CNR, sez. Trieste  - Trieste,
ITALY\\ $^{\sharp}$Istituto di Struttura della Materia, CNR-Area della
Ricerca di Tor Vergata - Roma, ITALY}


\maketitle

\begin{abstract}
We have investigated the electronic and magnetic structures of
Sr$_2$FeMoO$_6$ employing site-specific direct probes, namely
x-ray absorption spectroscopy with linearly and circularly
polarized photons. In contrast to some previous suggestions, the
results clearly establish that Fe is in the formal trivalent state
in this compound. With the help of circularly polarized light, it
is unambiguously shown that the moment at the Mo sites is below
the limit of detection ($< 0.25~\mu_B$), resolving a previous
controversy. We also show that the decrease of the observed moment
in magnetization measurements from the theoretically expected
value is driven by the presence of mis-site disorder between Fe
and Mo sites.

\vspace*{1cm}

PACS number(s): 75.25.+z, 75.30.Vn, 75.20.Hr

\end{abstract}


The observation of colossal magnetoresistance (CMR) in the perovskite
mixed valent manganites has led to a renewed interest in ferromagnetic oxides.
It is believed that the double exchange mechanism in presence of
strong electron-phonon couplings arising from Jahn-Teller
distortions is responsible for the observed properties in the
manganites~\cite{millis}. Recently, double perovskite Sr$_2$FeMoO$_6$
was established as a new CMR
material~\cite{koba}. This compound, in contrast to the
manganites, has certain technologically desirable properties, such
as a substantial MR at a low applied field even at the room
temperature. From a fundamental point of view, it is even more important
to note that crystallographic data does not indicate any
JT distortion and the lattice does not appear to play any
significant role in this compound. Furthermore, the system is an undoped
one in contrast to the manganites. Thus, Sr$_2$FeMoO$_6$ is in
principle a simpler system to understand its properties in
detailed theoretical terms. Inspite of this apparent simplicity,
there are surprisingly many open issues concerning the basic electronic
and magnetic structures of this compound.

In this compound, each of the Fe$^{3+}$ ($S$ = 5/2) and Mo$^{5+}$
($S$ = 1/2) sublattices are believed to be arranged
ferromagnetically, while the two sublattices are coupled to each
other antiferromagnetically. It has been suggested~\cite{koba}
that the system is in a half metallic ferrimagnetic (HMFM) state
leading to its fascinating CMR properties. However, there appears
to be several controversies concerning the real nature of this
compound. One neutron diffraction study~\cite{neutron} reported
the absence of any moment at the Mo sites, suggesting Mo to be
essentially nonmagnetic, whereas another study~\cite{mori}
suggested $\sim 1~\mu_B$ at each Mo site. Moreover, analysis of
M\"{o}ssbauer results have been interpreted both in terms of
Fe$^{3+}$~\cite{nak69,ssc} and Fe$^{2.5+}$~\cite{linden} states.
Thus, it is obvious that even the basic issues concerning the
electronic and magnetic structures of this compound have not been
settled so far. Since the analysis of neutron and M\"{o}ssbauer
data are model dependent, it is obviously necessary to obtain
site-specific direct information concerning the electronic and
magnetic properties of this compound. Additionally, in the
originally proposed magnetic structure~\cite{koba}, the system is
expected to have a moment of 4 $\mu_B$ per formula unit (f.u.) due
to the ferrimagnetic coupling between the Fe$^{3+}$ 3$d^5$ and
Mo$^{5+}$ 4$d^1$ configurations. However, the observed saturation
magnetization ({\it M$_S$}) is often found~\cite{koba,tokura,our}
to be about 3~$\mu_B$/f.u. The origin of this reduction in the
magnetic moment is also not clear at present. We address all these
issues combining linear and circularly polarized x-ray absorption
spectroscopy (XAS), with its ability to provide direct,
site-specific electronic and magnetic informations. In addition to
providing the magnetic structure of this compound and explaining
the reduction in the observed moment, our results also suggest
that this compound cannot be described within the conventional
double exchange mechanism.

Sr$_2$FeMoO$_6$ can exist with varying extent of mis-site disorder
between the Fe and Mo sublattices. Synthesis and characterization
of highly ordered ($\sim$ 90\%) and extensively disordered
(ordering of $\sim$ 30\%) Sr$_2$FeMoO$_6$ have been described in
our earlier publication~\cite{ssc}. The experiments were carried
out at the 4.2R circularly polarized beamline at Elettra
Synchrotron Radiation Facility. The measurements were performed at
77 K, which is well below the magnetic ordering temperature
($\sim$ 420 K)~\cite{nak68}. The sample surface was cleaned {\it
in situ} by scraping with a diamond file. The degree of circular
polarization at the relevant photon energy was approximately 85\%.

In order to address the valence state of Fe in such oxides, it
is most suitable to probe the Fe 2$p_{3/2}$~($L_3$) XAS which
exhibits very clear differences
between formal Fe$^{2+}$ and Fe$^{3+}$ states. Specifically, the
2$p_{3/2}$ absorption edge of all Fe$^{2+}$ species in an octahedral
crystal field exhibits a main peak at a lower energy, followed by
a weaker peak or shoulder at a higher energy. The ordering of the
peaks is reversed for Fe$^{3+}$ species~\cite{Laan} providing an easy
way to identify the formal Fe valence state independent of the extent
of covalency. We have recorded a high
resolution ($\sim$ 0.3 eV) Fe 2$p_{3/2}$ absorption spectrum of
Sr$_2$FeMoO$_6$ with linearly polarized light (Fig.~1). From this
figure, it is evident, even in absence of any detailed analysis that
{\em only a Fe$^{3+}$ valence state} is consistent with
the experimental result, exhibiting a weaker lower energy shoulder
and a higher energy main peak. However, in order to provide a
quantitative description  of the spectral features and, more importantly,
of the ground state, it is
important to carry out detailed calculations including
hybridization effects with the ligands within a cluster model on an
equal footing as core-valence and valence-valence multiplet
interactions and crystal-field effects, as
the participation of the ligand levels in determining the spectral
features may be significant~\cite{dd}. In order to minimize the number of
free parameters, we fix the multiplet interaction strengths, $F^2_{dd}$
(9.7~eV), $F^4_{dd}$ (6.1~eV), $F^2_{pd}$ (5.4~eV), $G^1_{pd}$
(3.9~eV), and $G^3_{pd}$ (2.2~eV) to 80\% of the atomic
Hartree-Fock values to account for the solid state screening.
Additionally, the hopping parameter strengths, $pd\sigma$,
$pd\pi$, and $pp\sigma$ of -1.7, 0.9 and 0.45 eV, respectively,
are guided by a tight-binding fitting~\cite{priya} of the
spin-polarized {\it ab initio} band dispersions. Moreover, we fix the
multiplet averaged 2$p$ core~-~3$d$ valence Coulomb interaction
strength, $U_{pd}$, to be 1.2 times that of $U_{dd}$ between the
3$d$ electrons, according to the usual practice. Thus, we are left
with only two adjustable parameters, namely $U_{dd}$ and the
O~2$p$~-~Fe~3$d$ charge transfer energy, $\Delta$. Since the resulting
many-body problem within a complete basis approach~\cite{dd}
involves nearly 30,000 basis states, the calculations were carried
out within the Lanczos algorithm. We obtain a remarkably good
description of the spectral features with  $U_{dd}$~=~4~eV  and
$\Delta$~=~3~eV for the Fe$^{3+}$ configuration, as shown in
Fig.~1.  $U_{dd}$ and $\Delta$ estimated here are consistent with
those in other octahedral Fe$^{3+}$ oxides. The many-body ground
state has 60.2\% $d^5$, 34.5\% $d^6$$\underline{L}^1$ and
5.1\% $d^7$$\underline{L}^2$ character, suggesting the system to
be somewhat more ionic than even LaFeO$_3$~\cite{ashis}. It should be noted
here that it was not possible to describe the spectral features at
all starting with a formal Fe$^{2+}$ configuration and then including
configuration interaction for any choice of parameter
strengths, conclusively establishing the 3+ valence state of Fe in
Sr$_2$FeMoO$_6$.

While the XAS with linearly polarized light at the Fe $2p$ edge
provides a definitive description of the site-specific electronic
structure, it is not as specific to the magnetic structure as
x-ray magnetic circular dichroism (XMCD) results would be. In
order to specifically investigate the magnetic structure, we have
carried out XAS at the Fe~$2p$ and Mo~$3p$ edges with circularly
polarized light. We use a lower resolution ($\sim$ 1.0 eV at the
Fe 2$p$ edge) to improve substantially the signal-to-noise ratio,
though this smoothens out the detailed spectral features which are
very similar to those with linearly polarized light (Fig. 1). In
Fig.~2(a), we show the photon-flux normalized
polarization-dependent Fe 2{\it p} XAS spectra, $\mu^+$ and
$\mu^-$ for highly ordered Sr$_2$FeMoO$_6$, corresponding to the
helicity parallel and anti-parallel to the Fe 3{\it d}
majority-spin direction, respectively. The XMCD spectra,
$\triangle$$\mu$~=~$\mu^+$ - $\mu^-$, also shown in the same
panel, clearly shows a substantial magnetic signal, indicative of
a large Fe moment. The corresponding experimental $\mu^+$, $\mu^-$
and $\triangle$$\mu$ spectra for Mo 3{\it p} edge are shown in
Fig.~2(b). It is evident from the XMCD spectrum of Mo that any
magnetic moment at these sites is below the detection limit ($<
0.25~\mu_B$). The present result is in clear contradiction with
the suggestion of a measurable moment ($\sim 1~\mu_B$) at the Mo
sites~\cite{koba,mori}, while it is in agreement with a previous
neutron diffraction measurement~\cite{neutron} where no moment
could be detected at the Mo sites.

The above results, in conjunction with already known properties of
Sr$_2$FeMoO$_6$, provide an understanding of the magnetic
structure. The XMCD spectra establish a large moment at the Fe
site, while negating the possibility of a substantial moment at
the Mo sites. However, the electronic structure of this compound
with a formal Fe$^{3+}$ state requires the existence of another
electron, nominally associated with a Mo$^{5+}$ 4$d^1$
configuration and being responsible for the metallic behavior. Our
results establish that the spin density arising from this single
electron is not substantially at the Mo site. Since this electron
is delocalized, it is not unreasonable to expect that the
wavefunction of the electron will be spread over several sites.
Band structure results, based on spin-polarized LMTO-GGA
calculations~\cite{koba,tanu} in fact clearly show that the states
at the $E_F$ are almost equally contributed by Fe~3$d$, Mo~4$d$
and O~2$p$ states, suggesting an average of $\sim$~0.3~${\mu_B}$
down-spin density at each Fe, Mo and six oxygen sites. The present
experimental result at the Mo~3$p$ edge suggets that the
spin-density is in reality further reduced ($< 0.25~\mu_B$) at the
Mo sites, compared to the single particle calculations. Thus, it
appears that the FeO$_6$ octahedron carry more than 0.75~${\mu_B}$
down-spin density, rather than $\sim$~0.6~${\mu_B}$ suggested by
the band structure results. The suggestion of a substantial
down-spin moment contribution at the Fe site is supported by our
many-body cluster calculations (see Fig.~1), where the ground
state wavefunction is found to have an average down-spin $d$
occupancy of 0.45, somewhat larger than that suggested by the band
structure results. Combining all these evidences, it would appear
that the delocalized electron spin density is transferred from the
minority spin of the Mo-sites {\it via} hybridization to
Fe~($\sim$~45\%) and O~($\ge$~30\%) with less than about 25\% of
spin-density at the Mo site, thereby spreading over several sites.
Thus, it appears that the delocalized spin density,
antiferromagnetically coupled to the localized up spins at the Fe
sites, prefers to be spatially closer to the central Fe sites,
thereby gaining a stronger antiferromagnetic coupling~\cite{prl}
between the localized and the delocalized spins rather than
residing at the farther Mo sites. While the double exchange
mechanism, applicable to the manganites, has often been invoked to
describe these ordered double perovskite systems, the present
results clearly suggest a new physics for this class of compounds
compared to manganites. In the DE mechanism, the localized spin at
the Mn site and the delocalized electrons, largely residing at the
same atomic site, are coupled {\em ferromagnetically}. In
contrast, the present system is reminiscent of previously
discussed Zhang-Rice singlet formation in the context of high
T$_C$ cuprates~\cite{zhang}. In that case the localized moment at
the central Cu site is coupled antiferromagnetically with the
doped delocalized spin-density spread over the central Cu and the
nearest neighbor oxygen sites to form a singlet state. In the
present case, the localized Fe {\it S}~=~5/2 state couples
antiferromagnetically with the spin density of delocalized {\it
S}~=~1/2 state to form a {\it S}~=~2 state.

Having established the basic magnetic structure of
Sr$_2$FeMoO$_6$, we now address the issue of consistently
observing a lower $M_S$ value for this compound than is expected
on the basis of the simple ionic picture, in all reported results.
In this context, it is important to note that Sr$_2$FeMoO$_6$
always appears with a finite concentration of mis-site disorder
where a pair of Fe and Mo exchange their crystallographic
positions. The best compounds have been reported to have
$\sim$~90\% ordering of the Fe and the Mo sites~\cite{koba,ssc}.
In order to investigate whether such a mis-site disorder can be
responsible for the observed reduction of the moment from the
ideal value of 4 to about 3 $\mu_B$ and whether the reduction in
the total magnetization is related to a corresponding loss in the
local magnetic moment of Fe, we have also recorded the XAS at the
Fe $2p$ edge of the extensively disordered Sr$_2$FeMoO$_6$ with
circularly polarized light. These edges along with the XMCD result
at Fe 2{\it p} edges for the disordered sample are shown in
Fig.~3(a), while in Fig.~3(b), we compare the XMCD signals from
the ordered and disordered Sr$_2$FeMoO$_6$. It is evident from the
spectra that the magnetic moment on individual Fe ions decreases
remarkably with decreasing ordering. In order to quantify our
results, we have calculated the orbital, spin and total moments at
the Fe sites from these spectra with the assumption of negligible
magnetic-dipole moment, using the well established sum
rules~\cite{thole,teramura}. All the individual spin, orbital and
total moments of these two samples are shown in Table 1. In these
cases, we find that $m_{orb}$ is very small, due to the
approximately $3d^5$ configuration of Fe ions. The resulting total
moments estimated from the XMCD signals are 1.68 and 1.36
$\mu_B$/Fe for the ordered and the disordered systems,
respectively. It is to be noted that the magnetic moments obtained
from the XMCD results are considerably smaller than the total
magnetic moments obtained from bulk magnetization
measurements~\cite{koba,our}. Such discrepancies are well known in
the literature~\cite{fuji}, and may arise from many factors. It
has been variously attributed to uncertainties in data analysis,
limitations of the applicability of the atomic sum rules arising
from non-ideal geometry in real experiments and/or solid-state
effects, and non-saturation of the magnetization at modest
magnetic fields near the surface region. Thus, the absolute value
of the moment estimated from the XMCD results is {\it per se} not
a useful quantity, though the extensive XMCD literature shows that
relative changes in the magnetic moments estimated from XMCD is a
very reliable quantity. For our purpose, we first establish this
point explicitly by comparing the XMCD results (Fig.~3 inset) of
the ordered Sr$_2$FeMoO$_6$ with a bulk magnetic moment of
$\sim$~2.81~${\mu_B}$/f.u. at 77~K and a closely related compound,
Sr$_2$FeMo$_{0.3}$W$_{0.7}$O$_6$, with a bulk magnetic moment of
3.64~${\mu_B}$/f.u. at 77~K. The XMCD results (Table~1) measured
at the same temperature clearly suggest an approximately
42$\pm$2\% drop in the XMCD moment compared to the bulk one for
{\it both} the samples. Thus, having established the efficacy of
probing the relative changes of magnetization in these and related
systems employing XMCD, our results on the ordered and disordered
Sr$_2$FeMoO$_6$, shown in Fig. ~3 and Table 1, clearly establish a
remarkable decrease in the magnetic moment at the Fe sites with
increasing mis-site disorder. These experimental results are also
consistent with the recent band structure calculations~\cite{tanu}
within a supercell for mis-site disorders between the Fe and Mo
occupancies. The band structure results suggest that a complete
disorder would result in a 34\% decrease in the moment at the Fe
sites, while a 50\% order would have a 21\% decrease of the moment
compared to the fully ordered sample. The present experimental
result of a 23\% decrease in the Fe moment for the ``disordered"
sample with about 30\% ordering is consistent with these band
structure results. Thus, it appears that the decrease in the
magnetic moment invariably observed for the so-called ordered
Sr$_2$FeMoO$_6$ is mainly due to the presence of finite ($\sim
10\%$) mis-site disorder. The origin of the decrease in the Fe
moment in presence of mis-site disorder is essentially due to the
destruction of the half-metallic ferromagnetic state of the
fully-ordered ideal system, thereby transferring $d$-electrons
from the up-spin to the down-spin bands~\cite{tanu}.

In summary, site-specific x-ray absorption spectroscopy with
linearly polarized light established the formal valency of Fe in
Sr$_2$FeMoO$_6$ to be 3+. Detailed investigation of x-ray magnetic
circular dichroism data confirms a large moment at the Fe site.
Our results provide a direct evidence for a negligible ($<
0.25~\mu_B$) magnetic moment at the Mo site, thereby suggesting
that the delocalized electron spin density, coupled
antiferromagnetically to the localized Fe-spins, is delocalized
over several sites including the neighboring FeO$_6$ octahedra and
indicating a novel origin of magnetism, different from the
conventional double exchange mechanism. A comparison of XMCD
results from the ordered and the disordered samples establishes
that the presence of mis-site disorder between the Fe and Mo sites
even in the so-called ordered samples is responsible for the
observed drop in the magnetic moment from the expected value of
4~$\mu_B$/f.u. to experimentally observed value of about
3~$\mu_B$/f.u.

This project is supported by DST, Government of India and Italian
Ministry of Science under the program of cooperation in Science
and Technology. We thank C. Carbone and P. Mahadevan for useful
discussions.

\begin{table}

Table~1.~~~Spin moments, Orbital moments and Total moments in $\mu_B$/Fe at 77 K.\\

\begin{tabular}{c c c c}
Compound & $M_{spin}$ & $M_{orb}$ & $M_{tot}$
\\ \hline Ordered Sr$_2$FeMoO$_6$  & 1.71 & -3.6$\times$ 10$^{-2}$ & 1.68
\\ Disordered Sr$_2$FeMoO$_6$ & 1.44 & -7.6$\times$ 10$^{-2}$ & 1.36
\\ Sr$_2$FeMo$_{0.3}$W$_{0.7}$O$_6$  & 2.15 & -7.3$\times$ 10$^{-2}$ & 2.07

\end{tabular}
\end{table}

\begin{center}
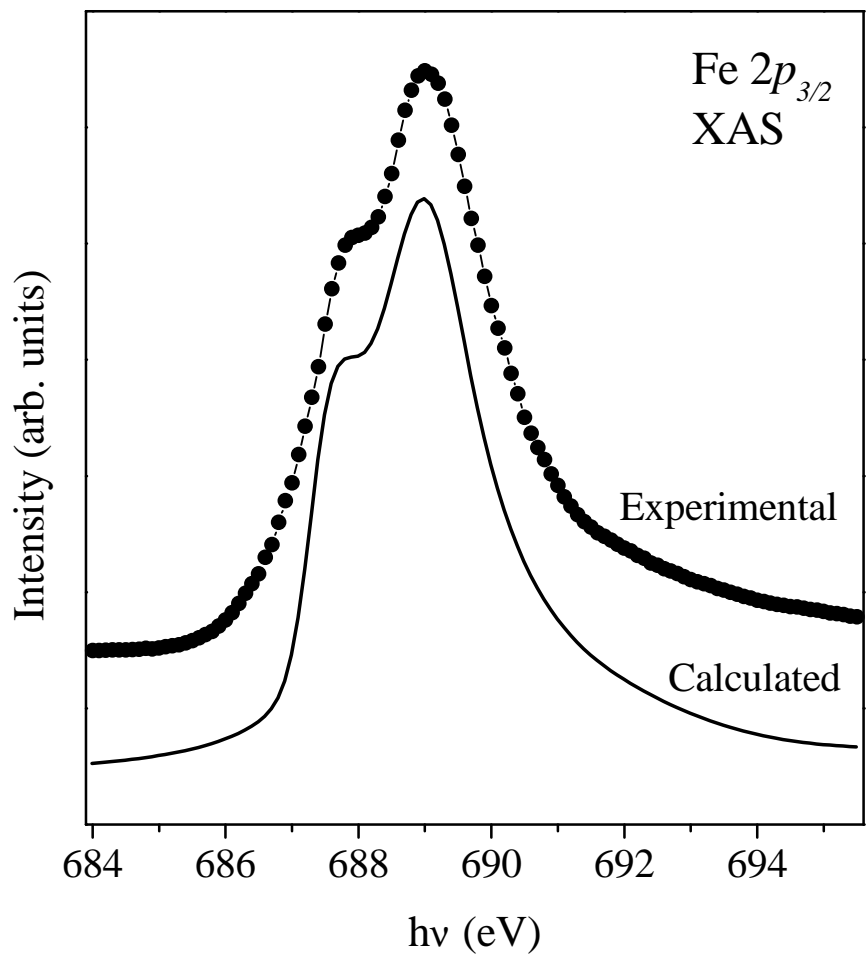
\begin{figure}
\caption{Experimental and calculated Fe 2$p_{3/2}$ x-ray
absorption spectrum for ordered Sr$_2$FeMoO$_6$.}
\end{figure}
\end{center}

\begin{center}
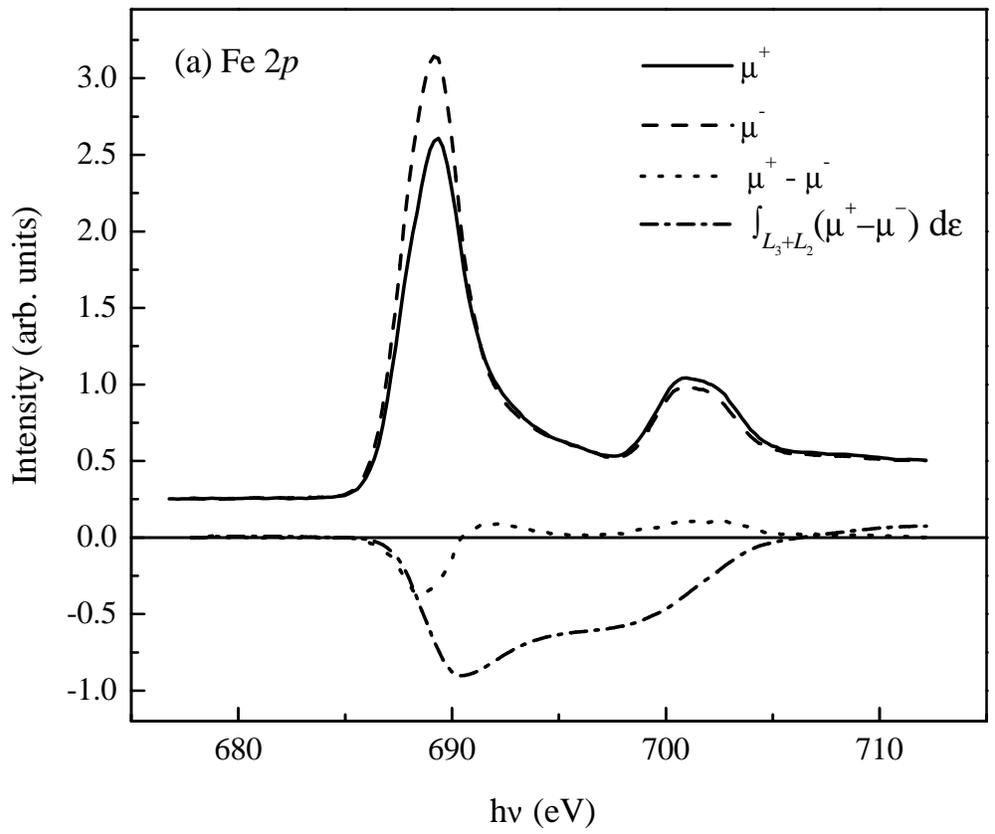
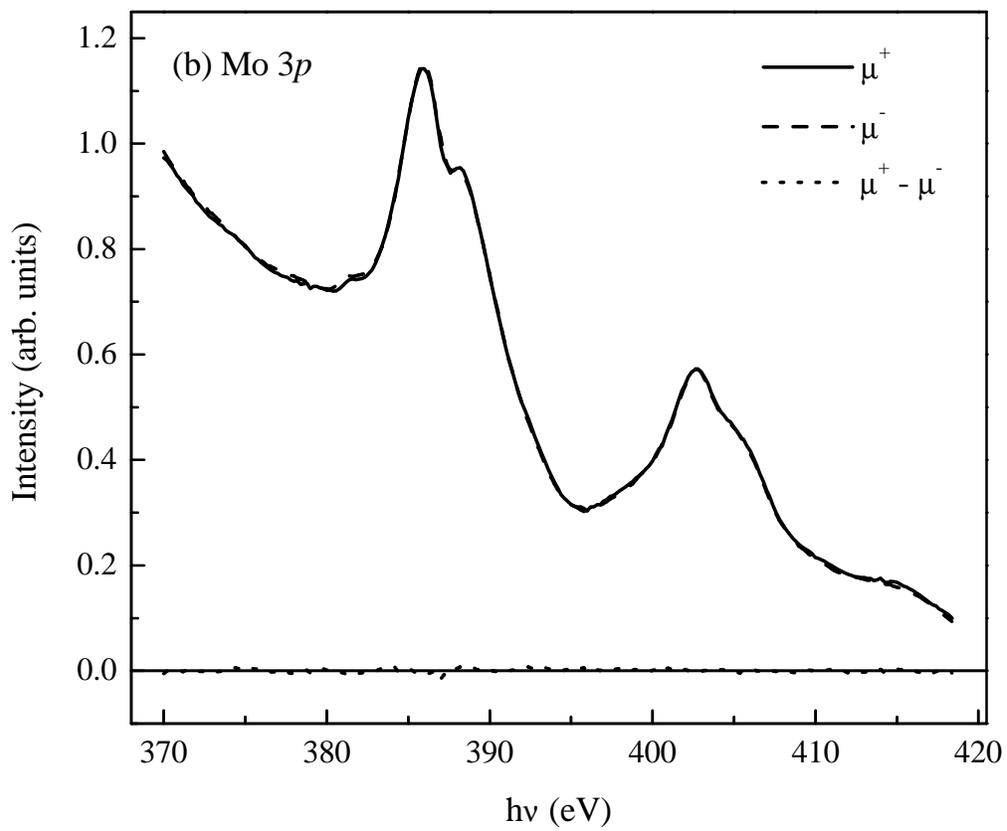
\begin{figure}
 \caption{x-ray absorption spectra at (a)~Fe~2$p$, and (b)~Mo~3$p$
edges for ordered Sr$_2$FeMoO$_6$, measured using circularly polarized light.
In the lower panels, circular dichroism signals, the
difference between the absorption for right and left circularly
polarized light at these edges are shown. The integral difference spectrum for
Fe~2$p$
edge is also shown in the lower panel of (a).}
\end{figure}
\end{center}

\begin{center}
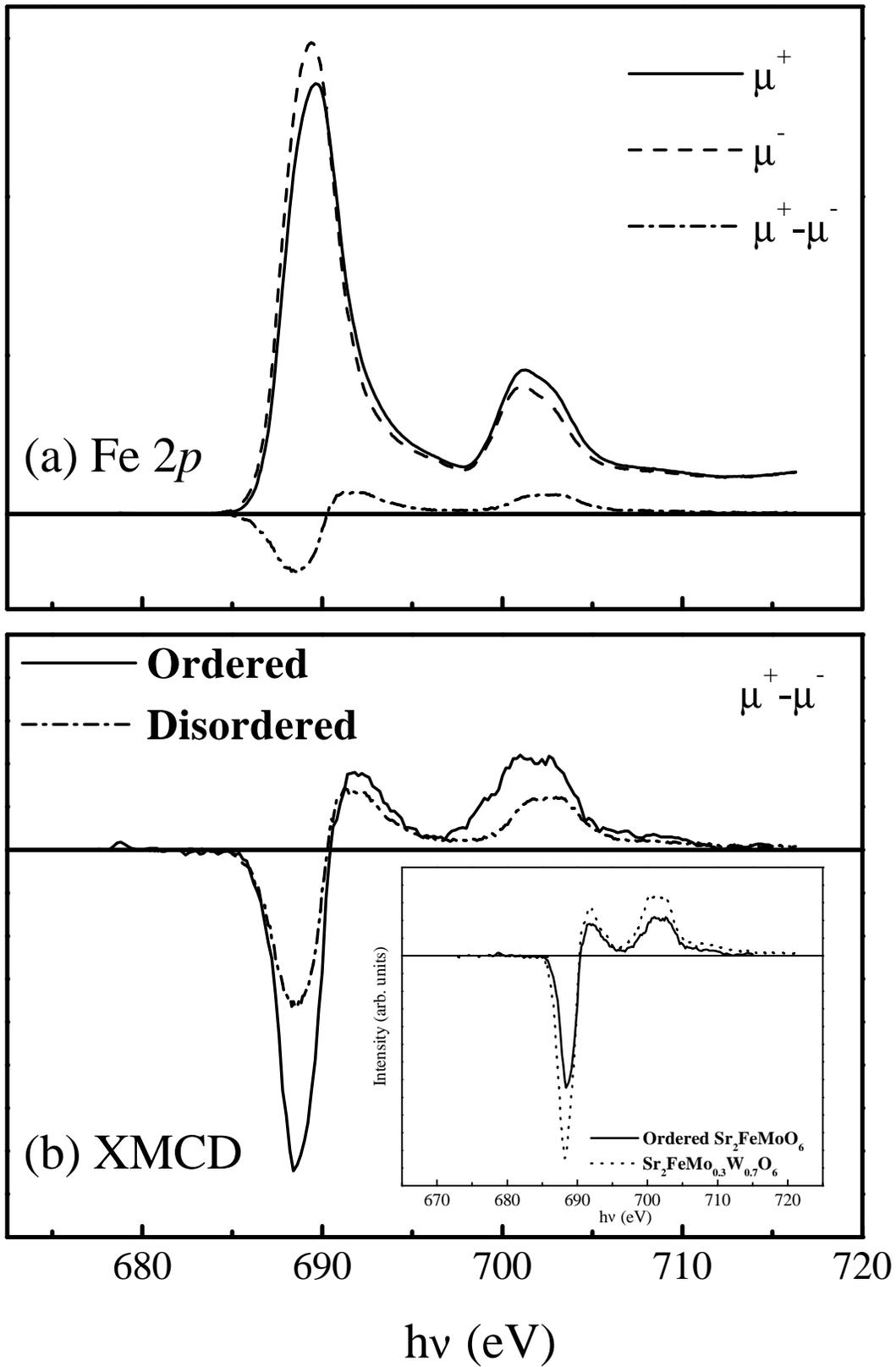
\begin{figure}
\caption{(a) x-ray absorption spectra at Fe~2$p$-edge for
disordered Sr$_2$FeMoO$_6$, measured using circularly
polarized light and the corresponding XMCD signal. (b) XMCD signals for the
ordered
and disordered Sr$_2$FeMoO$_6$. In the inset, the XMCD signals
for ordered Sr$_2$FeMoO$_6$ and Sr$_2$FeMo$_{0.3}$W$_{0.7}$O$_6$
are shown.}
\end{figure}
\end{center}

\end{document}